# III-Nitride Nanowire Based Light Emitting Diodes on Carbon Paper

Michael A. Mastro[1], Travis J. Anderson[1], Marko J. Tadjer[1], Francis J. Kub[1], Jennifer K. Hite[1], Jihyun Kim[2], Charles R. Eddy, Jr.[1]

[1] US Naval Research Laboratory, 4555 Overlook Ave. SW, Washington, DC 20375 USA
[2] Department of Chemical and Biological Engineering, Korea University, Seoul, South Korea

**Abstract**
This article presents the use of flexible carbon substrates for the growth of III-nitride nanowire light emitters. A dense packing of gallium nitride nanowires were grown on a carbon paper substrate. The nanowires grew predominantly along the a-plane direction, normal to the local surface of the carbon paper. Strong photo- and electro-luminescence was observed from InGaN quantum well light emitting diode nanowires.

**Introduction**
The III-nitride light emitting diode (LED) is a critical component in several modern technologies such as general white light illumination and liquid crystal displays. [1-6] To service these technologies, there has been a constant push towards increased light output per watt of energy supplied with a secondary optimization of efficiency per square area. [7-14] The standard fabrication utilizes an expense laden process to build a multi-layer thin-film structure on rigid substrates such as sapphire. [15-26]

A contrasting application would be a low-emittance device covering a non-planar architectural structure, e.g., a pillar, or flexible element. To make such as device economically and mechanically feasible would require a low-cost process on a flexible substrate. This paper presents an approach to deposit functional light emitting diode as a multilayer core-shell nanowire structure via a vapor-liquid-solid (VLS) growth mechanism on a flexible carbon paper substrate.

**Experimental** The general fabrication of the nanowire consisted of growth of an n-type GaN:Si core as observable in Figure 1. The was followed by the formation of an InGaN-well/GaN:Mg shell around the core. The thickness of the InGaN shell layer is approximately 5 nm and, the outer thickess of the GaN:Mg layer is 200 nm after 5 min of growth. The thickness of the InGaN/GaN:Mg sheath was directly proportional to growth time. The structure of the nano-wires was designed [4] to create a thin, single InGaN well for quantum confinement of the injected carriers, and a thicker GaN:Si core and GaN:Mg sheath for optical confinement of the optical mode.

The nickel seeds required for VLS growth were formed from a 0.005 M nickel nitrate solution that was repeatedly dripped onto a carbon paper substrate and blown dry in $N_2$ before loading the carbon paper into a vertical impinging-flow MOCVD reactor. A 50-Torr, $N_2/H_2$ mixed atmosphere was used during the ramp to growth temperature. Trimethylgallium was flowed for 2 sec prior to the onset of $NH_3$ flow to prevent nitridation of the nickel seeds. The GaN nanowire core was grown at a temperature of 850ºC, a pressure of 50 Torr and a V/III ratio of 50. Under proper growth conditions, the metal catalyst particle captures reactants and enhances the growth rate perpendicular to the substrate, thus creating a pseudo-one-dimensional semiconductor wire. [5] The LED nanowires continued with growth of an InGaN quantum well shell at 600ºC and a V/III ratio of 250. [6] Thereafter, the GaN:Mg shell was immediately grown to avoid decomposition of the InGaN. The introduction of Mg dopant atoms is known to encourage a higher lateral growth rate relative to growth of undoped GaN nanowires under equivalent conditions. [7] The samples were cooled from growth temperature in pure nitrogen ambient to avoid rapid decomposition in hydrogen and, in the case of the samples with a p-type shell, to activate the acceptor dopant (Mg).

A backside Au contact provided current conduction to the coalesced n-type GaN layer via the highly conductive carbon paper substrate. A Ni/Au bilayer provided the top contact to the p-type GaN. Further description and properties of the carbon nanofoam paper are available at [27]. Structural characterization was performed with a LEO FE Scanning Electron Microscope (SEM). Photoluminescence measurements were carried out using a HeCd laser at 325 nm and an Ocean Optic QE6500 spectrometer. Direct-Current I-V characterization was performed using an HP4145 parameter analyzer. Electroluminescence (EL) images were taken using a standard digital camera mounted on the probe station.

**Results and Discussion** Images of the GaN nanowires on the carbon paper are observable in the scanning electron micrographs in Fig 1. The nanowires grow uniformly over the entire carbon paper surface. The growth of the GaN core is not entirely one-dimensional. A low yet observable in-plane growth leads to a slight taper and eventual coalescence at the base of the wire. The formation of a continuous layer is advantageous as it prevents a short circuit from the top electrode to the underlying carbon paper. In other words, under a vertical bias the current is forced to flow through the pn junction.

The direction of GaN growth is generally perpendicular to the local surface. Growth of III-nitride nanowires via a VLS mechanism under this growth condition tends to proceed in the <11-20> a-direction with an isosceles triangular cross-section exhibiting a distinct set of facets of (0001), (1-10-1), and (-110-1). [8]

The III-nitride wire length is proportional to growth time with 1 hr of growth yielding wires of approximately 10 μm in length. The III-nitride nanowires have an approximate 200 nm diameter over a length of several microns.

While the nanowires grow along the a-direction, a small amount of deposition directly on the carbon paper presents a range of diffracting planes at the nominal surface. A 2θ-θ X-ray diffraction pattern in Figure 2 presents sharp diffraction from the (11-20) planes as well as weaker diffraction from other planes.

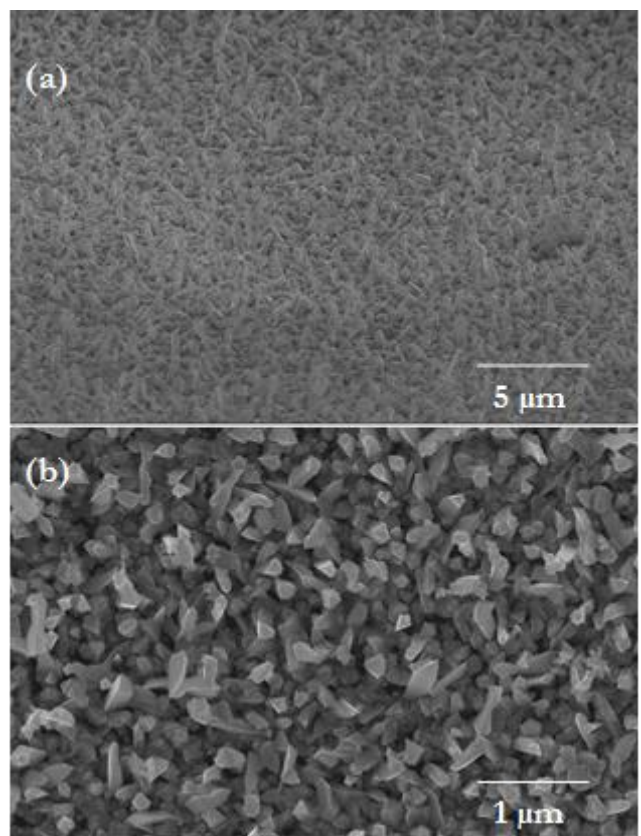

**Figure 1**. Electron micrograph of GaN nanowires on a carbon paper substrate at increasing level of magnification. A slight tapering is evident owing to growth at an elevated nanowire growth temperature although these conditions are known to produce higher quality GaN nanowires.

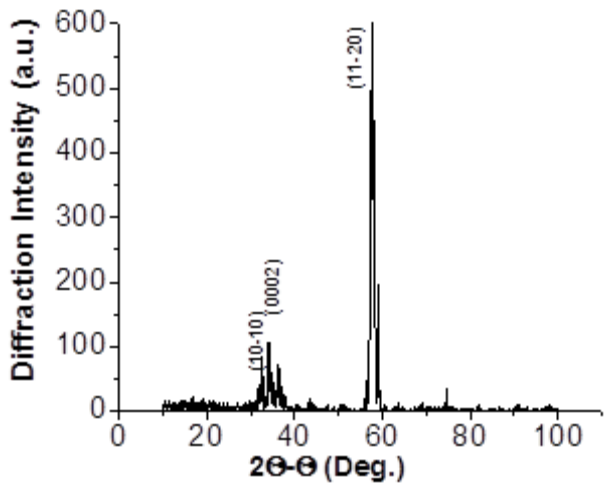

**Figure 2**. X-ray diffraction pattern of GaN nanowires on a carbon paper substrate. Some reactants impinging on the paper do not interact with the nickel catalyst and thus crystallize on the paper by uncontrolled nucleation.

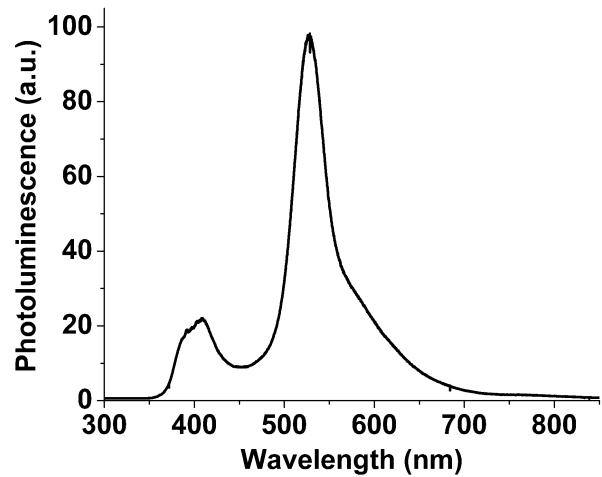

**Figure 3**. Photoluminescence spectrum of GaN:Mg / InGaN well / GaN:Si core-shell nanowires on a carbon paper substrate. The room temperature photoluminescence of the nanowires displays two major bands in the spectrum corresponding to the InGaN quantum well (540nm) and band-edge/ dopant-related luminescence transitions in the GaN (400nm).

Figure 3 shows the photoluminescence spectrum from GaN:Mg / InGaN well (sheath) / GaN:Si (core) nano-wires. The nano-wires display the expected GaN near band-edge dopant-based transitions (~400 nm) as well as the 530 nm InGaN luminescence from the quantum well. Additional experiments demonstrated that the InGaN peak varied with the composition of the InGaN as expected. Likely, the emission of the InGaN is slightly blue-shifted by the onset of quantum confinement in the 5-nm well. Broad-band yellow luminescence was present for growths at lower temperature and higher pressure. Yellow luminescence did not significantly contribute for nanowires formed in the growth conditions given above.

Figure 4 displays the current-voltage characteristics of a III-nitride nanowire-based light emitting structure on carbon paper. The carbon paper is highly conductive and effectively behaves as a metallic electrode. Electroluminescence was clearly visible by eye.

It is noteworthy that a persistence luminescence was observed for tens of seconds after removal of bias, either by removing the bias or physically removing the probes. Furthermore, mechanical deflection of entire sample was observed during bias and during the persistent luminescence phase. In an effort to understand this behavior, it is of note that carbon paper has an appreciable capacitance and that this trait is utilized to store charge in carbon paper based capacitors and super-capacitors. [9]

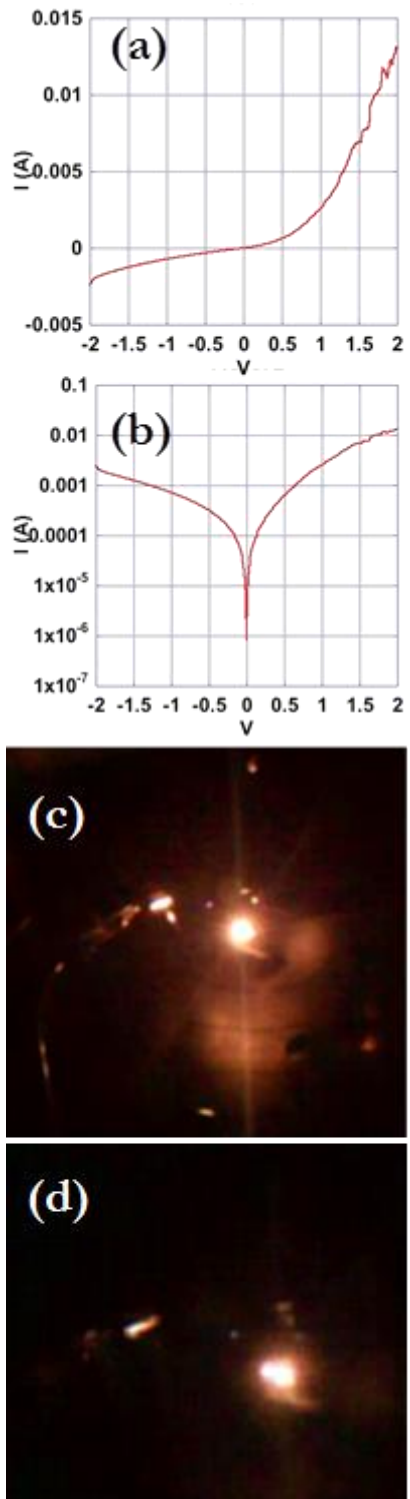

**Figure 4**. Electroluminescence intensity of GaN nanowires on a carbon paper substrate. (a, b) The current-voltage relation is characteristic of a functional pn junction with light emission (c) evident under bias. (d) Persistent luminescence for tens of seconds was observable after the bias was ended. Physical removal of the probes yielded a similar persistent luminescence.

It is known that GaN possesses large piezoelectric coefficients, and (similar to ZnO) GaN has been demonstrated as a piezo-electric energy harvesting element. [13] The exact mechanism and linkage of this apparent charge storage, mechanical deflection, and release as luminescence is currently under study.

**Summary** This investigation demonstrated high-quality, reproducible group-III nitride nanowire based light emitters on a conductive, low-cost, and flexible carbon paper substrates.

**Acknowledgements**
Research at the Naval Research Laboratory is supported by the Office of Naval Research